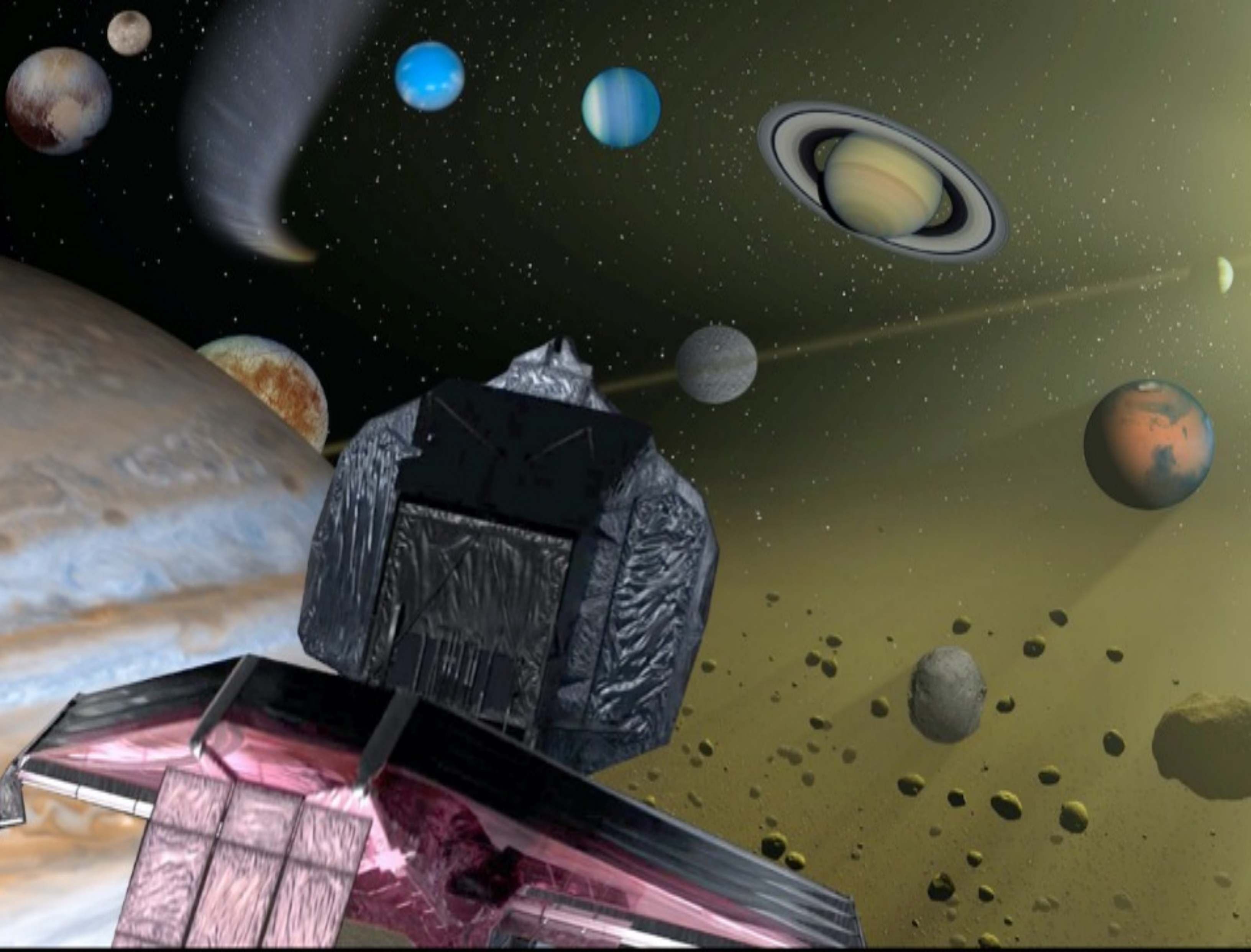

# A Lesson from the James Webb Space Telescope:

## Early Engagement with Future Astrophysics Great Observatories Maximizes their Solar System Science

***Heidi B. Hammel****, Primary Author*

Association of Universities for Research in Astronomy

Ph: 202-483-2101

Email: hbhammel@aura-astronomy.org

*with co-author* ***Stefanie N. Milam*** *from NASA/GSFC*

*Illustration on cover page: A fanciful montage shows the James Webb Space Telescope with the Solar System targets it will study with guaranteed time in Cycle 1.*

# A Lesson from the James Webb Space Telescope: Early Engagement with Future Astrophysics Great Observatories Maximizes their Solar System Science

*Primary author: Heidi B. Hammel (AURA)*

*Co-author: Stefanie N. Milam (NASA/GSFC)*

Endorsers of this white paper include:

Sushil Atreya
Tracy M. Becker
Jim Bell
Gordon Bjoraker
Paul K. Byrne *
Richard J. Cartwright
Thibault Cavalié
Jeff Cuzzi
Katherine de Kleer
Michael DiSanti
Edith C. Fayolle
Yan Fernandez
Estela Fernández-Valenzuela
Patrick Fry
Justin Garland
W.M. Grundy
Jennifer Hanley
Dean C. Hines
Bryan Holler
Michael Kelley
Laszlo Kestay (Keszthelyi)
Zhong-Yi Lin
Timothy A. Livengood
Adam McKay
Sarah E. Moran
Michael "Migo" Mueller *
Thomas Müller *
Max Mutchler
Conor Nixon
Michael C. Nolan
Glenn Orton
Andrew S. Rivkin
Michael T. Roman
Tony Roman
Abigail Rymer
Britney Schmidt
Agustin Sanchez-Lavega
Pablo Santos-Sanz
James Sinclair
Michael Sitko
Cristina Thomas
Matthew Tiscareno
Anne Verbiscer
Geronimo Villanueva
Chick Woodward
Padma Yanamandra-Fisher

** Endorsement received after submission to Planetary Decadal*

## Abstract

Astrophysics facilities have been of tremendous importance for planetary science. The flagship space observatory Hubble Space Telescope has produced ground-breaking Solar System science, but when launched it did not even have the capability to track moving targets. The next astrophysics flagship mission, the James Webb Space Telescope, included Solar System scientists in its science team from the earliest days, with the result that Webb will launch with a diverse program and capabilities for Solar System exploration. The New Great Observatories, as well as future ground-based facilities, offer the opportunity for a robust suite of observations that will complement, enhance, and enable future Solar System exploration. We encourage the Planetary Science and Astrobiology Decadal Survey 2023-2032 (hereafter "Planetary Decadal") to overtly acknowledge the prospects for excellent Solar System science with the next generation of astrophysics facilities. We hope the Planetary Decadal will further encourage these missions to continue to formally involve Solar System scientists in the science working groups and development teams.

## Introduction

There is a rich history of Solar System science using assets initially developed for deep-sky astrophysics.  Such observations complement *in situ* planetary missions in important ways: they provide preliminary assessment of bodies prior to missions; they add capabilities that complement those of *in situ* missions; they generate large-scale context data for *in situ* missions; they have long timelines to provide broader perspectives on time-variable phenomena; they can identify mission targets for future spacecraft exploration; and more.

A pair of recent review articles highlights science from across the Solar System from the Spitzer Space Telescope, which did engage with the Solar System science community from an early stage of development. Lisse et al. (2020) and Trilling et al. (2020) review this engagement, and also discuss contributions to the science of comets, Centaurs, and Kuiper Belt Objects (Lisse et al. 2020), and outline advances in our knowledge of asteroids, planets, and the Zodiacal cloud (Trilling et al. 2020).

Hubble Space Telescope, in contrast, did not engage deeply with the Solar System community throughout its development years, and consequently had a rocky start for Solar System observations: when launched, Hubble was not able to even track moving targets. Through heroic efforts of several individuals, this was rectified. The Hubble Space Telescope has gone on to revolutionize many areas of Solar System research, including (and this is by no means a comprehensive list): evidence for Europa plumes (Roth et al. 2014); studies of the Kuiper Belt and distant comets (e.g., Fraser and Brown 2012; Li et al. 2020); context imaging and spectra for Mars missions (Bell and Ansty 2007); assessing the stability of atmospheric features on giant planets (e.g., Hueso et al. 2020; Hsu et al. 2019); imaging aurorae on giant planets (Yao et al. 2019); characterizing the ocean interiors of giant planet satellites (Saur et al. 2015); and discovering moons and a follow-up target for the New Horizons missions to Pluto and beyond (Porter et al. 2018).

Mission planners for the "Next Generation Space Telescope (NGST)" were determined to avoid the Hubble mistake and hence took a page from the Spitzer playbook: in 2003, when interdisciplinary scientists were competitively selected by NASA for the NGST Science Working Group, two of the six were card-carrying Solar System scientists (Jonathan Lunine and Heidi Hammel). This mission, subsequently renamed James Webb Space Telescope, has benefited from thousands of hours of engagement with Solar System scientists over the past twenty years (Fig. 1). [1]

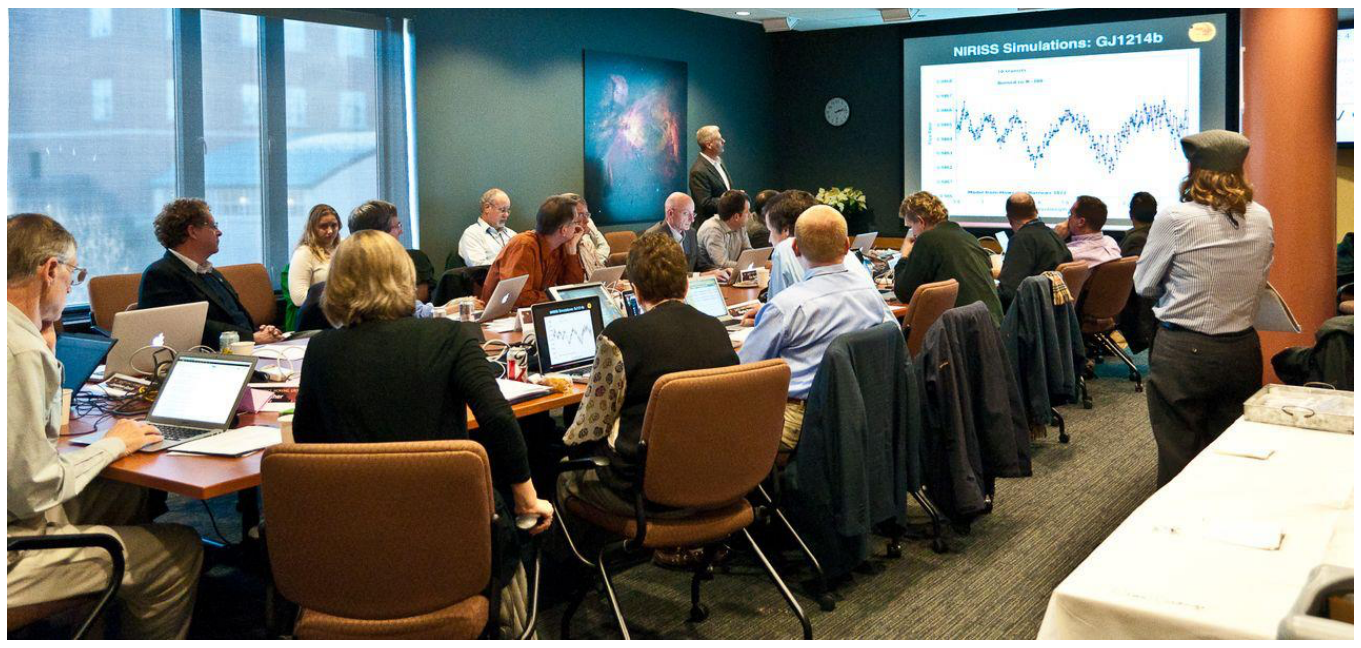

*Fig. 1. One of dozens of Webb Science Working Group meetings over the past two decades. White paper lead H. B. Hammel is in the foreground (in black). In addition, the SWG has held a weekly telecon every Monday for close to twenty years (excluding holidays and weeks when the group met in person).*

## Benefits

The benefit to early and frequent engagement with the mission development teams are numerous. Solar System scientists have a seat at the table throughout the decades of planning, development, and construction. Moving target tracking capabilities can be properly defined and implemented. Operational modes that benefit Solar System observations can be crafted and incorporated. Webb's ability to observe bright objects via sub-array readouts, or the ability to detect faint objects near bright objects through careful stray-light analyses, are driven in part by Solar System needs.

A program's interaction with the broader Solar System community is far more effective when a team exists of well-trained and deeply-engaged planetary scientists. The Webb program has hosted many years of workshops at DPS and LPSC meetings. The Webb team has also led planning workshops specifically geared for Solar System science.

[1] This estimate is based on attendance at weekly science meetings and participation in several multi-day science working group meetings each year over the past two decades; multiple workshops and presentations at planetary science conferences in recent years; many more hours of writing, presentations, etc.; and, finally, significant work by the PIs of the programs listed in Table 1.

**Table 1: Cycle 1 Solar System Observations with James Webb Space Telescope ***

| Lead(s) | Program | Hours | Object |
|---|---|---|---|
| Andrew Rivkin | 1244 | 9.4 | Large Asteroids and Trojan Asteroids |
| Cristina Thomas | 1245 | 9.8 | Near-Earth Objects |
| Leigh Fletcher | 1246 | 4.8 | Jupiter's Great Red Spot |
| Leigh Fletcher | 1247 | 12.0 | Saturn, including rings and small moons |
| Leigh Fletcher | 1248 | 9.3 | Uranus |
| Leigh Fletcher | 1249 | 4.2 | Neptune |
| Geronimo Villanueva | 1250 | 9.0 | Sub-surface oceans of Europa and Enceladus |
| Conor Nixon | 1251 | 12.8 | Titan: Climate, Composition, and Clouds |
| Michael Kelley | 1252 | 13.6 | Spectral mapping of a comet's inner coma |
| Stefanie Milam | 1253 | 5.0 | Target of Opportunity Comet |
| Pablo Santos-Sanz | 1271 | 2.0 | Target of Opportunity TNOs stellar occultations |
| Geronimo Villanueva | 1415 | 5.1 | Mars |
| Stefanie Milam | 1255 | 2.1 | Hammel part of JWST Medium-Deep Fields |
| *John Stansberry, Aurelie Guilbert-Lepoutre, Alex Parker, Dean Hines, Jonathan Lunine* | *1191, 1231, 1254, 1272, 1273* | *81.1* | *Various TNOs and similar objects including Pluto/Charon, Eris, Orcus, Makemake, Varuna, Triton, Sedna, Chariklo, Haumea, Quaoar, Chiron, Gonggong, and others *** |
| *Imke de Pater and Thierry Fouchet* | *1373* | *33.1* | *Jupiter System Early Release Science †* |
| **TOTAL** | | **213.3** | |

* All programs except the last two rows are part of Hammel's Guaranteed Time Observations and are available for immediate access as well as available to proposals for archival research, with zero proprietary time. ** Proprietary time was required for TNO observations because they are shared with other Guaranteed Time Observers/Teams. † Jupiter System observations are an Early Release Science program led by Drs. Imke de Pater and Thierry Fouchet.

A consequence of this engagement with the Solar System community is that Webb will field a robust set of Guaranteed Time Observations (GTO) and Early Release Science (ERS) programs for Solar System Objects, see Table 1 as well as the papers in a special issue of PASP devoted to JWST Solar System Science (PASP 2016).

## Future missions

NASA's "New Great Observatories" refer to four flagship concept studies conducted for the US Astro2020 decadal survey activity (Fig. 2). All four space telescope concepts – nicknamed LUVOIR, Lynx, Origins, HabEx – included Solar System observations in their science plans. As was true for Hubble and Spitzer, and will be true for Webb, for each of these facilities there will be unique ground-breaking science results to complement future *in situ* Solar System space missions (see *greatobservatories.org* for the complete study reports for all four missions). The first gate for these missions is a top ranking in the Astro2020 decadal survey; thereafter, discussions with interested parties will be in earnest.

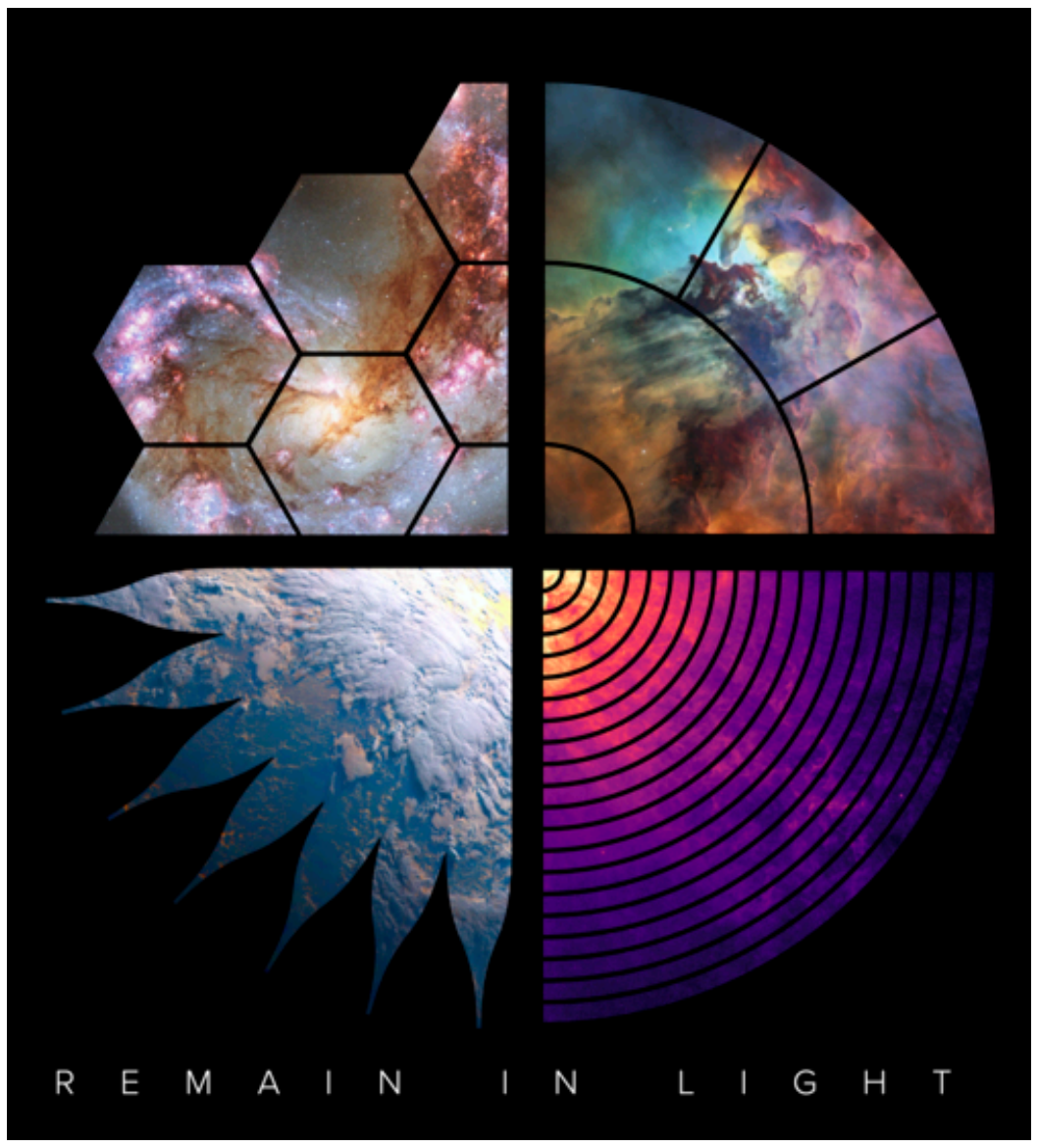

*Fig. 2. The New Great Observatories are concepts for future astrophysics missions; all have capabilities of interest to Solar System scientists.*

On the ground-based side, Keck Observatory, Gemini Observatory, the VLT, and other facilities continue to provide remarkable Solar System science. At longer wavelengths, ALMA has already achieved significant planetary science advances. Astronomers are already contemplating what the future may hold. The Giant Magellan Telescope (GMT, first light 2029) and the Thirty Meter Telescope (TMT, first light 2027) are working collaboratively with the US-Extremely Large Telescope Program (US-ELTP) to create to telescopes of similar class to the European Extremely Large Telescope (E-ELT, first light in 2025), with facilities in each hemisphere for full-sky coverage. These observatories offer the promise of exquisite Solar System science with their expected high spatial resolution and high sensitivity. TMT has included teams of planetary scientists in their science planning[2]; GMT has considered some applications[3]; and E-ELT touches lightly on the topic.[4] The proposed next-generation VLA would deliver milliarcsecond resolution. For all of these facilities, there is much room for future exploration of Solar System capability.

[2] For TMT, see https://www.tmt.org/page/our-solar-system and https://www.tmt.org/download/Document/36/original

Beyond the optical ELTs is uncharted territory. Whether the traditional path is followed to the increased aperture of an "overwhelmingly large telescope," or whether a new trail is blazed into optical interferometry, there will undoubtedly be use cases for Solar System observations. Perhaps they will map surface spectroscopy of Kuiper Belt objects and outer Solar System moons, or find evidence for activity at cometary nuclei at extreme distances from the Sun, or uncover the source of time variability of giant planet ring systems. The quest for higher resolution at radio wavelengths may lead to space interferometer arrays. These advanced facilities will offer the promise of new vistas in planetary exploration.

## Conclusion

Astrophysics facilities have for generations aided in our quest to understanding the formation and evolution of objects within our own Solar System. The immediate future continues that tradition with robust programs for planetary science. Looking to the future, we envision a continued collaboration between these communities to enable continued Solar System exploration and advancement.

Solar System science is maximized when planetary scientists are engaged with missions from earliest stages. NASA's New Great Observatory teams were attentive to this and that is laudable. We now need the commitment to be carried through as these missions move forward. With new technologies, sensitivities, and capabilities will come new challenges for the community to overcome and enable the next generation of remote planetary science.

*We encourage the Planetary Decadal Survey to explicitly recognize the importance of continued engagement of space astrophysics facilities with the Solar System science community, and to also explicitly encourage NASA to follow the precedents with Spitzer and Webb: ask the Astrophysics leadership within headquarters to engage directly with the Solar System science community from the very earliest stages of the mission development.*

*We also ask the Planetary Decadal Survey to encourage current and future large ground-based facilities to continue to broaden and strengthen their engagement with*

[3] For GMT, see https://www.gmto.org/Resources/GMT-ID-01464-Chapter_3_GMT_Science_Case.pdf
[4] For E-ELT, see https://www.eso.org/public/about-eso/faq/faq-elt/#29

*the Solar System science community. Such engagement is greatly facilitated by the hiring of active Solar System researchers to observatory staff positions.*

*Finally, we urge the Planetary Decadal Survey to explicitly encourage Solar System scientists themselves to participate in the Science Working Groups and science teams for future astrophysics facilities both on the ground and in space. Early engagement throughout the development and construction phases ensures the best opportunities to maximize the science output for Solar System observations.*

## References

Bell, J. F., and T. M. Ansty (2007). Icarus 191, pp. 581-602, DOI: 10.1016/j.icarus.2007.05.019

Fraser, W., and M. E. Brown (2012). Astrophysical Journal 749, Issue 1, id.33, DOI 10.1088/0004-637X/749/1/33

Hueso, R. et al. (2020). Icarus, Volume 336, article id. 113429, DOI: 10.1016/j.icarus.2019.113429

Hsu, A. I. et al. (2019). Astronomical Journal, Volume 157, Issue 4, article id. 152, DOI: 10.3847/1538-3881/ab0747

Li, J. et al. (2020). Astronomical Journal, Volume 159, Issue 5, id.209, DOI: 10.3847/1538-3881/ab7faf

Lisse, C. et al. (2020). Nature Astronomy (under revision)

Milam, S. N. et al. (2016). Publications of the Astronomical Society of the Pacific, Volume 128, Issue 959, DOI: 10.1088/1538-3873/128/959/018001

PASP (2016): https://iopscience.iop.org/issue/1538-3873/128/959

Porter, S. B. et al. (2018). Astronomical Journal, Volume 156, Issue 1, id.20, DOI: 10.3847/1538-3881/aac2e1

Roth L. et al. (2014). Science 343, pp. 171-174, DOI: 10.1126/science.1247051

Saur, J. et al. (2015). Journal of Geophysical Research: Space Physics, Volume 120, Issue 3, pp. 1715-1737, DOI: 10.1002/2014JA020778